\documentclass[conference]{IEEEtran}
\IEEEoverridecommandlockouts
\usepackage{cite}
\usepackage{amsmath,amssymb,amsfonts}
\usepackage{algorithm}
\usepackage{algpseudocode}
\usepackage{graphicx}
\usepackage{textcomp}
\usepackage{xcolor}
\usepackage{longtable}
\usepackage{hyperref}
\usepackage{multirow}

\makeatletter

\def\ps@IEEEtitlepagestyle{%
  \def\@oddfoot{\mycopyrightnotice}%
  \def\@evenfoot{}%
}
\def\mycopyrightnotice{%
   {\footnotesize 979-8-3503-7240-3/24/\$31.00 \copyright 2024 IEEE\hfill}
}

\def\BibTeX{{\rm B\kern-.05em{\sc i\kern-.025em b}\kern-.08em
    T\kern-.1667em\lower.7ex\hbox{E}\kern-.125emX}}

    \newcommand{\hdk}[1]{\textcolor{black}{#1}}
    
    \definecolor{wjcolor}{rgb}{0.6, 0.4, 0.8}
    \newcommand{\wj}[1]{\textcolor{black}{#1}}

\begin{document}

\title{Optimal Planning of Electric Vehicle Charging Stations: Integrating Public Charging Networks and Transportation Congestion
}


\author{\IEEEauthorblockN{Jingbo Wang, \textit{Graduate Student Member, IEEE,} Harshal D. Kaushik, \textit{Member, IEEE,} \\Jie Zhang, \textit{Senior Member, IEEE}}
    \IEEEauthorblockA{\IEEEauthorrefmark{1}\textit{The University of Texas at Dallas,  Richardson, Texas, U.S.} \\
    \{jingbo.wang, harshal.kaushik, jiezhang\}@utdallas.edu}
}

\maketitle

\begin{abstract}

The \wj{adoption of} electric vehicles (EVs) represents a critical shift in personal mobility, \wj{fueled by policy support and advancements in automotive technology}. However, the expansion of EVs for long-distance travel is hindered by charging time concerns, the sparse distribution of charging stations, and the worsening waiting times due to congestion. The main objective of this work is two-fold: 1) first, to comprehensively analyze the existing public charging station robustness and effectively strategize for the new ones, and 2) secondly, to select the optimal chargers for long-distance journeys, by estimating the waiting time from current traffic congestion. This is achieved by accompanying effective EV charging strategies, pinpointing on the congestion points from the existing traffic, and the robustness of the current charging station infrastructure. Utilizing a real-time transportation and charging station dataset in Texas, we identify optimal charger placement strategies to minimize travel time by examining the congestion and charging time trade-offs. Our findings suggest that maximizing the constant current phase during charging enhances efficiency, crucial for long-distance travel. On the contrary, we also explore the negative impact of congestion on travel times and we conclude that sometimes it might be beneficial to exceed the constant current phase to avoid the congested charging stations.

\end{abstract}


\section{Introduction}

\hdk{A continuous} \wj{shift} to electric vehicles (EVs) \wj{represents} a pivotal \wj{transformation} in personal transportation, \wj{driven} by the \wj{pressing} need to \wj{lower} carbon emissions and \wj{decrease dependence} on fossil fuels. As EVs \wj{become the norm} for daily transportation, the critical challenge lies in enabling efficient long-distance EV travel, highlighting the importance of a comprehensive and accessible charging infrastructure. The proliferation of public charging stations is a positive trend, yet optimizing their distribution is key to enhancing the robustness and reach of the charging network for extended journeys.

This paper examines the interplay between the availability of public chargers with their respective waiting times according to current traffic, \hdk{considering} the characteristics of EV batteries, and strategies for minimizing their charging periods. \wj{Determining the optimal placement and capacity of charging stations is important for enhancing user experience while minimizing overall charger investment costs.} 
\hdk{Authors in} \cite{EV_planning_10} introduce a reliability index for allocating EV charging stations within a 33-bus distribution network.
\hdk{Location planning of EV charging stations have been studied in \cite{EV_planning_14, EV_planning_15, EV_planning_16}, considering a generic traffic,} however, the existing studies have not accounted for real traffic networks. In this research, \hdk{we focus specifically on considering a real transportation network along with the existing real traffic flow there.} 

\hdk{A} Constant Current Constant Voltage (CCCV) technique is frequently employed for charging. Introduced by Shi et al. \cite{shi2022electric}, a pioneering remaining charging time estimation algorithm emphasizes the significance of the {constant current (CC)} phase to uphold high charging efficiency for long-haul travel.  Distinct EV models feature varied battery capacities and preferred charging State of Charge (SOC) ranges, affecting the determination of optimal charging site locations. For instance, the ideal recharging stops for a Tesla differ from those for a GM Bolt, even if the destination remains the same. 

Route planning involves finding route based on criteria such as driving distance, travel time, or energy consumption. Dijkstra's algorithm is widely used for \wj{finding the shortest distance in a network\cite{Dijkstra_1959},} although it can be slow for large graphs. Techniques \hdk{such as}  A* algorithm use heuristics for faster searches \cite{Hart_1968}. 

Route planning for electric vehicles \wj{becomes more complex} due to battery constraints, including limited range and \wj{energy} recuperation capabilities. This \wj{introduces the}  constrained shortest path problem \cite{Artmeier_2010}, where finding the shortest path considering battery constraints is essential. Modified versions of the Dijkstra's algorithm or A* can calculate all Pareto optimal paths, although this \wj{increases computational complexity.} Contractions hierarchies can also be used to speed-up multi-criteria path finding and to solve the shortest path in acceptable time \cite{Storandt_2012}.

 Baum et al. \cite{Baum_2017} proposed a method for solving the electric vehicle constrained shortest path problem, prioritizing paths that comply with battery constraints while minimizing travel time. Their approach suggests driving speeds and avoids the need to compute all Pareto optimal paths. They accelerate queries using a combination of contraction hierarchies and the A* algorithm. Razo and Jacobsen \cite{Razo_2016} proposed a model that relies on mobile network reservations to estimate charging station waiting times, aiding vehicle selection. 
Tan and Wang \cite{Tan_2015} utilized game theory to optimize travel costs for users and maximize revenues for charging stations, considering freely set hourly prices and grid overloading. \

To enhance practicality, our work in this veins estimates the shortest path by considering the actual traffic and real road network of the \hdk{Dallas Fort Worth (DFW)} area.  By utilizing real-time traffic data and congestion analysis, we can accurately forecast waiting times.
With the increase in EV adoption, a rise in vehicles on the road leads to inevitable congestion at charging sites, especially near major thoroughfares. This paper introduces an optimal charging strategy that accommodates congestion considerations and the efficiency trade-offs of extended CV phase duration to enhance total travel time. By aligning charging stations with driving ranges tied to optimal charging efficiency phases, the study aims to boost the practicality and appeal of EVs for long-distance travel.

Utilizing the Texas EV charging network as a case study, our analysis underscores the advantages of strategic charger placement and the impact of potential charging congestion along travel routes. This methodology significantly curtails overall charging times for long-distance travelers, encouraging wider EV adoption.


The rest of the paper is organized as follows. Section \ref{mieg} discusses our methodology in detail. It begins with \wj{a comprehensive analysis of current traffic conditions and the distribution of EV charging stations across Texas, identifying key areas with insufficient charging infrastructure, and provides recommendations for} future charger locations. Next, we \wj{introduce}  an algorithm to select appropriate charging stations  for minimizing the total travel time and charging duration. Section \ref{csft} delves into a case study assessing the performance of our algorithm using real dataset, integrating the densely populated Dallas-Fort Worth area with traffic congestion points, charging stations, and the actual road network. The paper concludes in Section \ref{conc}, summarizing the findings and their implications for future research and policy-making.


\section{Model For EV Charger Network Planning and EV Long-distance Planning}\label{mieg}


In this section, we outline our approaches for network planning and long-haul planning of EV travel. The first model pertains to network planning, aiming to identify areas with limited charger availability to serve congested regions. The second model outlines a methodology for planning long-haul EV travel.

We initiate by constructing a network that integrates the actual road network, existing EV chargers, and congestion points at various locations. Subsequently, we identify areas with scarce or nonexistent chargers to fulfill charging demands. These demands are estimated by drawing appropriate clusters around EV chargers.

Next, we estimate the waiting times at all charging stations within the area using available congestion points and charging station locations. This estimation primarily relies on the percentage of vehicles requiring charging, particularly EVs. By analyzing clusters around charging stations, we determine the percentage of vehicles heading to each station, aiding in calculating total waiting times. We then compute optimal routes considering both travel and waiting times, ensuring minimal charging time alongside travel time. 

Focusing on charger locations in Texas, sourced from the Department of Energy (DOE) database \cite{afdc}, our objective is to devise a customized charger network that identifies optimal charger locations for various EV models.
We utilize vehicle transportation dataset from the Texas transportation website {\cite{txdot_roadways}}, specifically annual average 24-hour vehicle counts across Texas, {with a current EV percentage of 1.4\% of the entire traffic {\cite{peng2024investigating}}}. 
The selection of optimal charger locations relies on a comprehensive understanding of travel distances, duration, congestion levels, and the Constant Current Constant Voltage (CCCV) charging method, specifically considering the lithium-ion battery's most efficient SOC window during the CC phase \cite{kostopoulos2020real}. 
\hdk{As EVs transition to the {constant voltage (CV)} phase, where voltage is steady but current diminishes, the overall charging power drops, prolonging charging duration and decreasing efficiency. Considering the diverse battery characteristics of EVs, the CC phase is associated with specific driving ranges, with the optimal charging point occurring when cell voltages reach their maximum.} 
Next, we elaborate on our methodology in detail. We commence by analyzing the current charging station network. Then, we discuss our strategy for deciding future charger locations. We end the section by proposing an algorithm to select charging locations for long-haul EV travel.

\subsection{Current Charger Network {Analysis}} \label{currentnetwork}

To enhance the EV charging network's efficiency and accessibility, we evaluated the current infrastructure, focusing on the range capabilities of EV models available as of 2023. Our findings indicate that most EVs designed for long-distance travel can journey between 209 to 353 miles on a single charge, with an average of around 281 miles { \cite{cnet_ev2023}}. However, for long-distance trips, the timing of charging stops—primarily during the {CC} phase to avoid the less efficient {CV} phase—reduces the effective travel distance to between 136 and 212 miles {\cite{kostopoulos2020real}}. This range, assuming {an SOC}  between 15\% and 80\%, is crucial for assessing the charging network's coverage. Based on this analysis, we ensure that within the charger network, the distance between consecutive nodes along the edges ranged from 136 to 212 miles. This strategy ensures that EVs can travel effectively within the CC charging phase, thus optimizing the total travel time.

\begin{figure}
    \centering
    \includegraphics[width=0.5\textwidth]{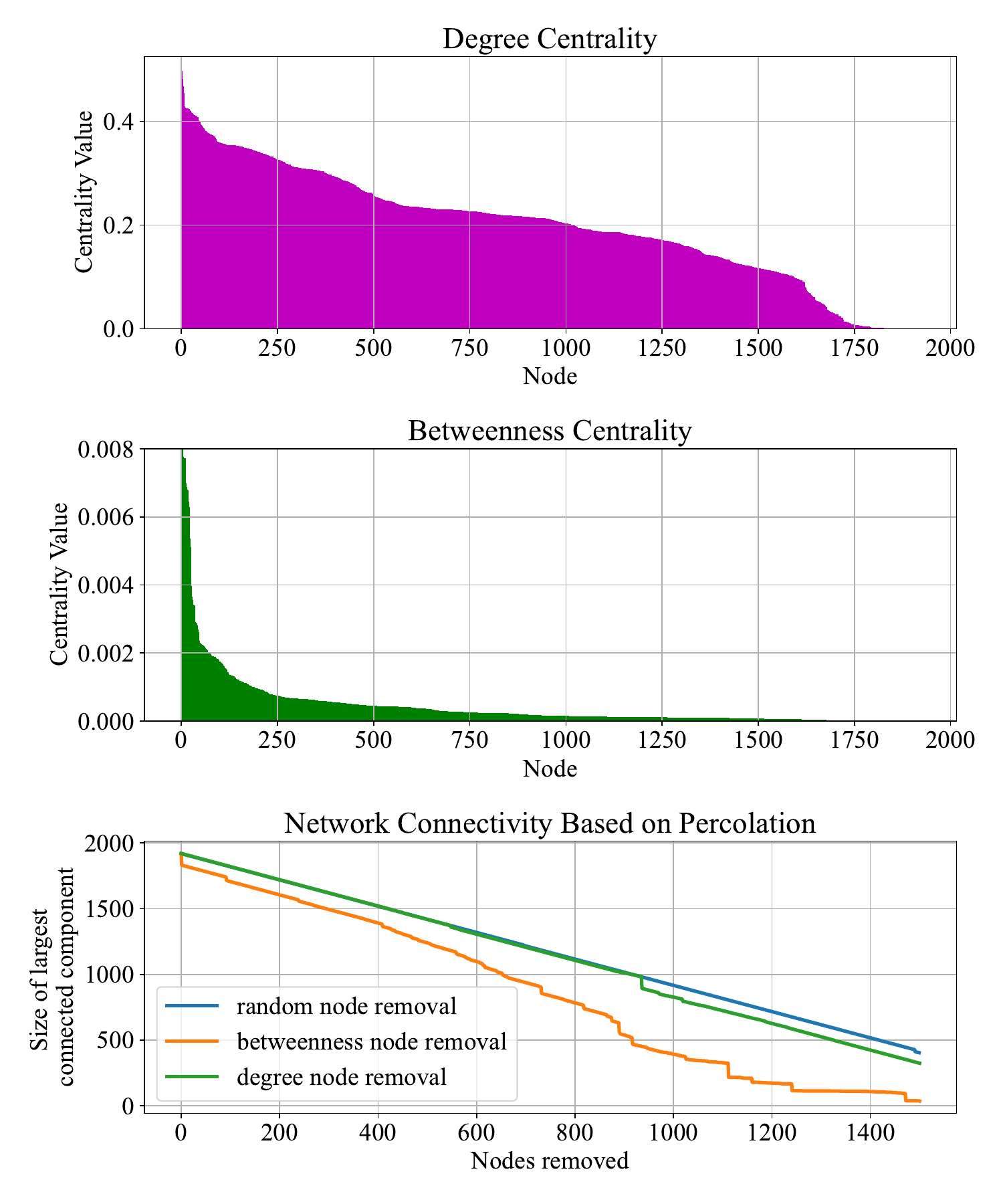}
    \caption{Current Texas public charger network robustness}
    \label{fig:centrality}
\end{figure}

Our review of the {current network state} utilized metrics {such as degree and betweenness centrality} to gauge station connectivity and percolation analysis {to evaluate the} resilience against disruptions, which is {presented in Fig. \ref{fig:centrality}}. Findings show a balanced network distribution, with the highest connectivity in {central Texas} and a gradual decrease in less {densely} populated areas, indicating a network aligned with population density and travel patterns.

Percolation analysis demonstrated the network's robustness over graphs \cite{zhang2023real} to random failures, but it also exposed vulnerabilities in the event of targeted attacks on critical nodes. This insight suggests enhancing critical nodes with redundancy and rapid-response repair capabilities to ensure continued network functionality.

\subsection{Optimizing Future Charger Network Development}
This section delves into an integration of the EV charger network data {within Texas} from \cite{afdc} and 24-hour vehicle counts at different locations in Texas \cite{txdot_roadways}. {We utilize} high vehicle counts as indicators of potential congestion zones. {From {\cite{shukla2019multi}}}, we established a 40-mile coverage radius for each charger, a conservative estimate that presumes EVs within this range would utilize the corresponding charger. Employing GeoPandas {\cite{kelsey_jordahl_2020_3946761} and a clustering technique}, we optimally cluster the area served in Texas by each charger.  As shown in {Fig. } \ref{fig:chargercoverage}, the green clusters represent the 40-mile radius coverage of each charger. The congestion points, plotted in red, serve as the demand points. The blue dots highlight areas currently not served by any available chargers, indicating a shortage of EV charging facilities in Texas and underscoring the need to plan for additional chargers. By using the K-Means clustering algorithm from scikit-learn, it aims to partition geographic locations into five clusters based on proximity, enabled us to strategically identify areas with the most significant need for new charging facilities. This approach is visualized by magenta dots in Fig. \ref{fig:chargerplanning}. We assess these clusters against traffic congestion data, which helps prioritize locations where new chargers can significantly alleviate congestion and improve the EV usage experience in Texas.



Incorporating these newly identified charging sites, we also propose an updated framework for route planning and charger selection for long-distance EV travel, aiming to enhance coverage and accessibility for EV users across the state {considering the real transportation network.

\begin{figure}
    \centering
    \includegraphics[width=0.4\textwidth]{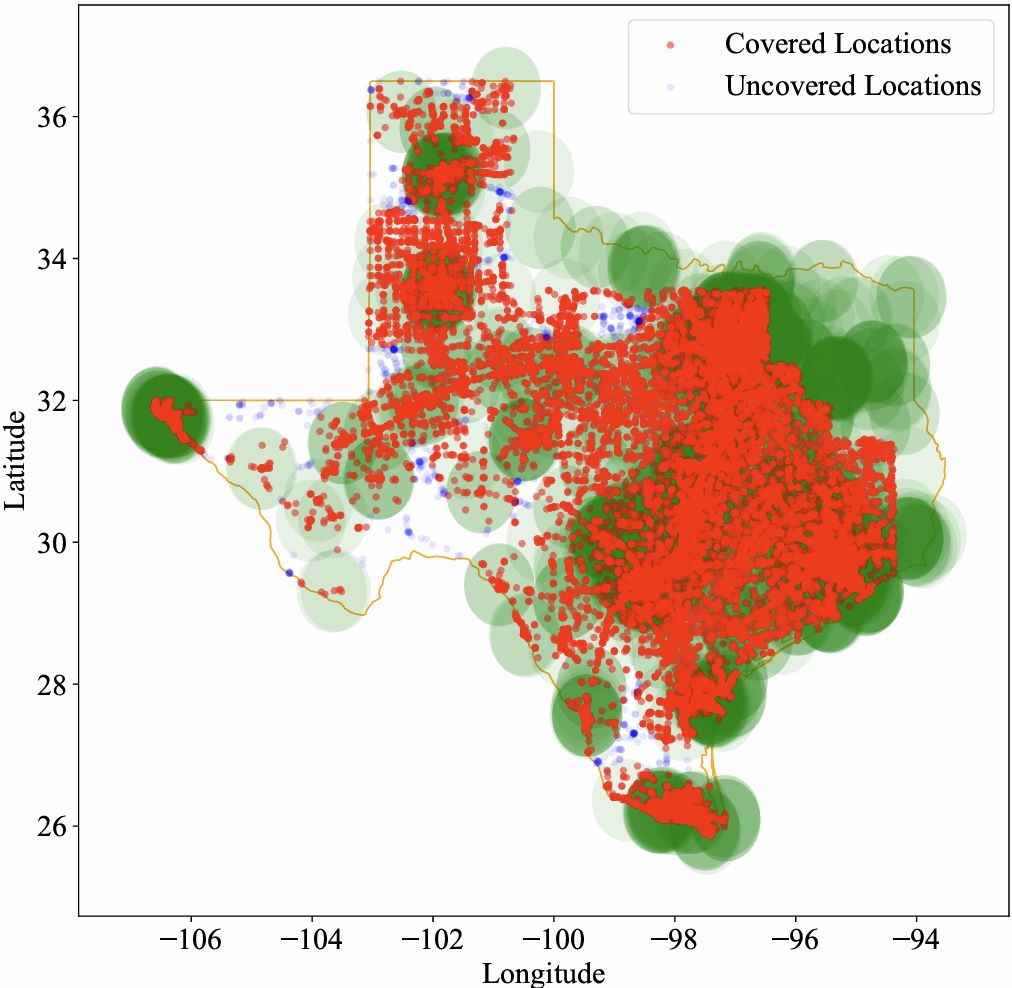}
    \caption{EV charger coverage in Texas as of 2023} 
    \label{fig:chargercoverage}
\end{figure}

\begin{figure}
    \centering
    \includegraphics[width=0.4\textwidth]{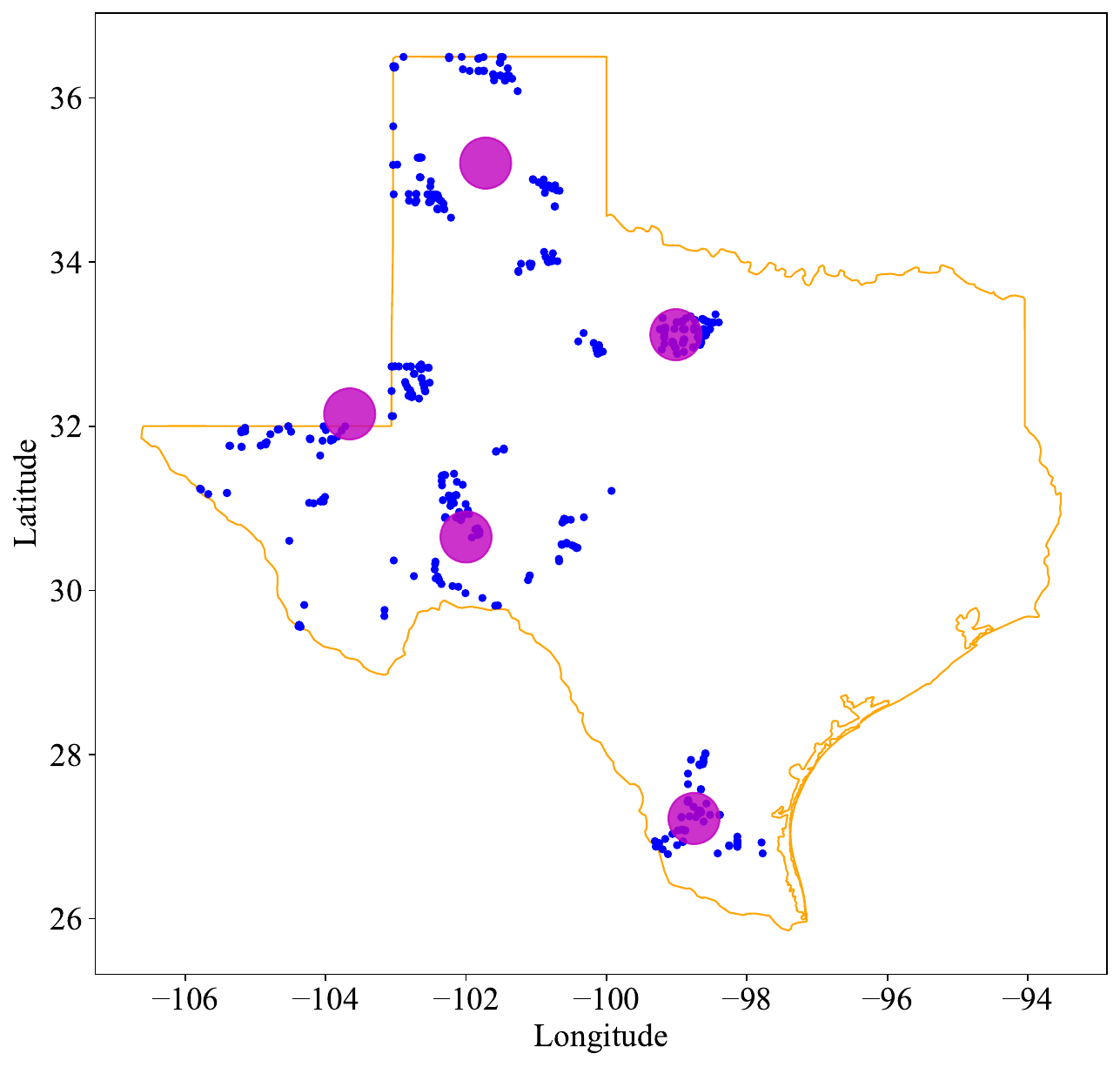}
    \caption{Congestion-driven EV charger planning}
    \label{fig:chargerplanning}
\end{figure}

\subsection{Selecting Prime Charging Spots for Long-Haul EV Travel}\label{spcs}
Despite significant reductions in charging times, EVs still require more time to charge compared to refueling conventional vehicles. Planning an EV journey thus necessitates careful consideration of charging station locations, with an emphasis on optimizing charging duration and minimizing potential congestion at these stations.

In addressing the challenge of congestion at charging stations, which could notably extend wait times, we introduce a predictive wait time model for each charging point. This model calculates wait times based on expected traffic flows and the projected number of EVs charging, thereby incorporating wait time into the total charging duration. In situations where congestion at stations is likely, our optimization algorithm may suggest alternate charging sites that, while possibly extending time in the Constant Voltage (CV) phase, provide a strategic balance to minimize overall wait times.

We formalize this as finding the shortest path problem. We start by defining a graph $G(V,E)$, where $V$ represents a collection of vertices and $E$ denotes a set of edges. For instance, consider the graph illustrated in Fig. \ref{fig:chargertradeoff}. The nodes, depicted in green, constitute the vertex set $V$, while the edges connecting them, displayed in grey, form the edge set $E$. The nodes representing charging stations are a subset of the vertex set $V$, denoted as $V_c$. For convenience, we define node set $N(\subset V)$ that includes all the nodes except the start and the end nodes. Subsequently, we introduce a binary decision variable $x_{ij}$ for each edge $(i,j)\in E$, where $x_{ij} = 1$ indicates that the edge $(i,j)$ is included in the shortest path. Now, from graph $G$, we have known the distance  to travel from node $i$ to $j$ (captured in matrix $C_{ij}$) and the waiting time at the charging station node $i\in V_c$ (captured in vector $W_i$). Using an appropriate multiplying factor $\alpha$, we define an objective function as in Eq. \eqref{eq:1} that corresponds to minimizing both travel distance and waiting time. Note that $y_i$ is an auxiliary binary decision variable, which takes the value of 1 when charging station $i \in V_c$ is visited.
\begin{align}
    \min  & \sum_{i,j\in E} c_{ij} x_{ij} +\alpha\sum_{i\in{V_c}}W_{i} y_i  \label{eq:1}\\
     & \sum_{i\in V} x_{ij} = \sum_{i\in V } x_{ji}  && \hspace{-3.6cm} \text{for }j\in N,\{(i,j),(j,i)\}\in E \label{eq:2}\\
     &\text{if } x_{ij} = 1 \ \Rightarrow \ y_i = 1 \text{ and }y_j = 1 \nonumber &&\\ & \hspace{1.8cm}  \text{ for }i/j\in V_c  \text{ and }(i,j)\in E && \label{eq:3}\\
     &c_{ij}x_{ij} \leq \lambda_{\text{threshold}} && \hspace{-1.8cm}\text{ for } (i,j) \in E \label{eq:4}\\
     & x_{ij} \in \{0,1\} &&\hspace{-1.8cm}\text{ for } (i,j) \in E\label{eq:5}\\
     & y_{i} \in \{0,1\} &&\hspace{-1.8cm}\text{ for } i \in V_c\label{eq:6}
\end{align}
Constraint \eqref{eq:2} ensures flow conservation at all nodes except the origin and destination nodes. Equation \eqref{eq:3} ensures that the auxiliary decision variable $y_i$ equals 1 at all visited charging station nodes $V_c$. Constraint \eqref{eq:4} ensures that the threshold value ($\lambda_{\text{threshold}}$) for the CC phase is not violated. Equations \eqref{eq:5} and \eqref{eq:6} ensure that the decision variables $x_{ij}$ and $y_i$ are binary, defined over the edge set $E$ and the node set $V_c$, respectively.

The approach adopted to address the shortest path problem, as outlined in the preceding formulation, is elaborated in Algorithm \ref{alg:optimal_charging}. This algorithm emphasizes the selection of optimal charging locations based on a holistic assessment of travel and charging times, as well as potential congestion impacts. 


\begin{algorithm}
\caption{Optimal Charging Route Planning}\label{alg:optimal_charging}
\begin{algorithmic}[1]
\State Define start and end locations for the route.
\State Initialize an empty list of potential charging {stations.}
\State Estimate charging {time w.r.t. battery and charger specification}.
\State Estimate potential waiting time due to {congestion.}
    \For{each origin and destination pair}
        \State Solve formulation for shortest path time.
        \State Calculate {Total Time (travel + charging  + waiting)}.
    \EndFor
\State Select an optimal set of chargers that minimize  {Total Time}.
\end{algorithmic}
\end{algorithm}

To better understand the actual waiting times, we analyzed vehicle flow data from the Texas transportation network, which provides average daily vehicle counts at measurement points over one year. Using the latest data from 2022, we converted these counts to hourly flow rates. Assuming that 1\% of these vehicles are EVs and that a similar proportion of EVs will require charging, we overlay charger locations onto these transportation points within a defined radius. Each charger's congestion time is estimated based on the assumption that each EV charges for 15 minutes, with a maximum wait time capped at one hour for each hourly interval.

The core of our optimization lies in the shortest path problem, tailored to the unique needs of the EV charging network. We aim to minimize total {travel and charging times}, acknowledging the constraints posed by EV range limitations, {charging duration, and congestion}. This is encapsulated in the formulation of the shortest path problem, as previously discussed. The approach we employed is detailed in Algorithm \ref{alg:optimal_charging}.
This approach not only optimizes the route based on physical distance but also ensures the battery is efficiently charged, minimizing overall travel time {by strategically reducing the time component in congestion}.


This refined strategy, which blends route efficiency with practical considerations of charger availability and traffic congestion, aims to provide an optimized travel experience for EV users. In the following section, we present a case study implementing Algorithm \ref{alg:optimal_charging} for planning EV travel within Texas.




\section{Case Study for EV Charger Network}\label{csft}
While our algorithm is capable of optimizing routes across the broader Texas road network for long-distance travel, we have chosen to validate its performance within the DFW metropolitan area due to the availability of detailed real road network data. We have customized the charger coverage distance  to 2 miles for the DFW area. 
This adjustment better reflects the typical urban driver’s willingness to deviate slightly from their regular route for charging, enhancing the relevance and application of our study to real-world urban scenarios.
The case study focuses on the implementation of a graph-based optimization algorithm, targeting the enhancement of EV charging infrastructure and the improvement of traffic management through advanced routing strategies. This section elaborates on the adaptations made to suit urban driving conditions and the resulting impacts on EV charger network efficiency and urban mobility. The existing charging stations along with their corresponding waiting time (at a particular day) is shown in Fig. \ref{fig:chargertradeoff}. For better visibility, we only show 10\% of the charging station waiting times.


Using the Python libraries NetworkX \cite{hagberg2020networkx} and OSMnx\cite{boeing2017osmnx}, we developed a detailed and realistic model of the DFW road network. This model incorporates comprehensive traffic data and geographic information, enabling dynamic adjustments to route planning based on current traffic conditions. The core of our approach is the development of a weighted graph, where nodes represent intersections or points of interest (such as EV charging stations), and edges symbolize the roadways connecting these points. Weights assigned to each edge correspond to travel times, which are adjusted dynamically based on data-driven predictions of traffic congestion and anticipated wait times at EV charging stations. This predictive capability is crucial for real-time traffic management, allowing the system to adapt routes on the fly to minimize delays and improve overall travel efficiency. The travel planning between two distant points, obtained by Algorithm \ref{alg:optimal_charging} is shown in Fig. \ref{fig:chargertradeoff_route}.

The integration of predictive analytics into our routing algorithm allows for advanced anticipation of traffic conditions and charger station loads, adjusting the travel routes in real time to optimize both travel time and energy usage. This method not only helps in managing the existing traffic and charger infrastructure more efficiently but also aids in planning future expansions of the EV charging network by identifying high-demand areas and potential congestion points.

By analyzing traffic flow patterns and charger usage statistics, the algorithm identifies optimal routes that balance short travel distances with minimal waiting times at chargers. This approach helps in significantly reducing the overall travel time for EV users, enhancing the attractiveness of EVs for urban residents, and supporting wider adoption.

The case study within the densely populated DFW metropolitan area validates the effectiveness of our algorithm, which innovatively incorporates congestion waiting times at charging stations--a factor often overlooked in the available methods in the literature. Specifically, our algorithm anticipates and mitigates potential delays caused by increased EV penetration and the resultant queuing at charging stations. The strategic inclusion of waiting times in the routing process not only refines travel time estimates but also  enhances overall trip efficiency. This foresight becomes increasingly valuable as EV adoption escalates, and the demand for accessible charging infrastructure intensifies. The evaluation of our algorithm's performance focuses on the reduction in total travel time in comparison to existing methodologies that do not account for charging station congestion.

\begin{figure}
    \centering
    \includegraphics[width=0.5\textwidth]{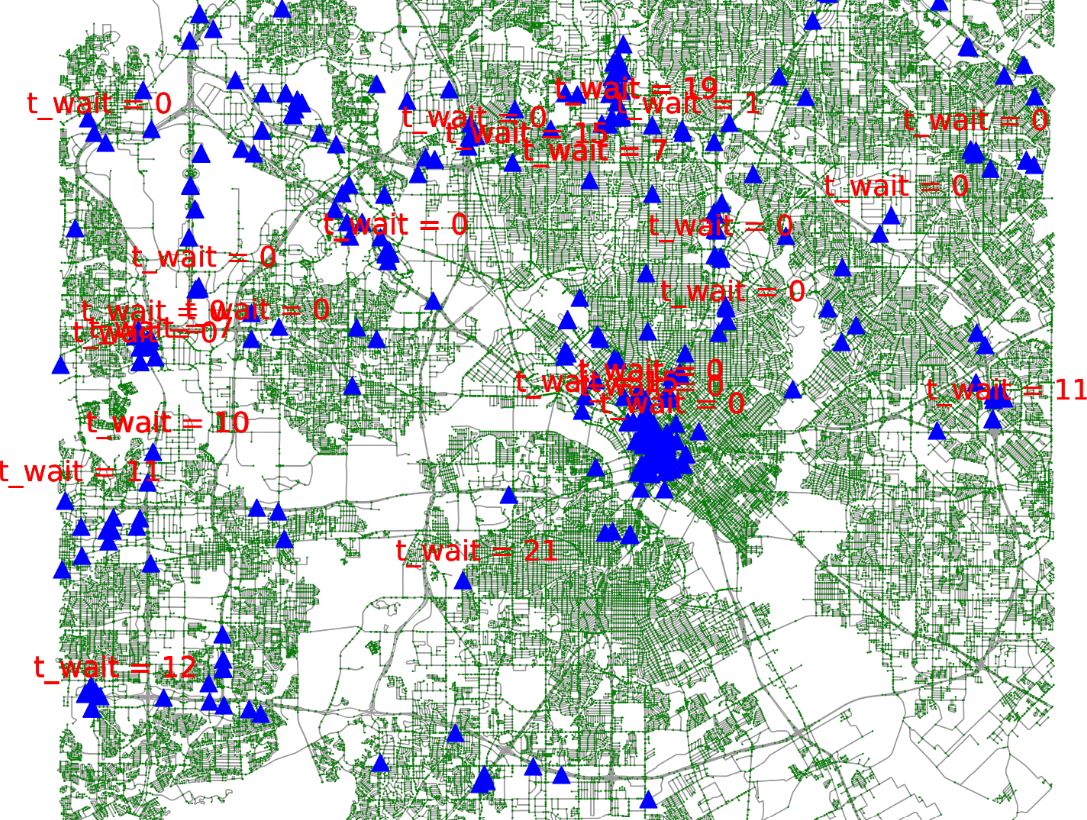}
    \caption{Waiting time due to congestion at each charger location}
    \label{fig:chargertradeoff}
\end{figure}

\begin{figure}
    \centering
    \includegraphics[width=0.5\textwidth]{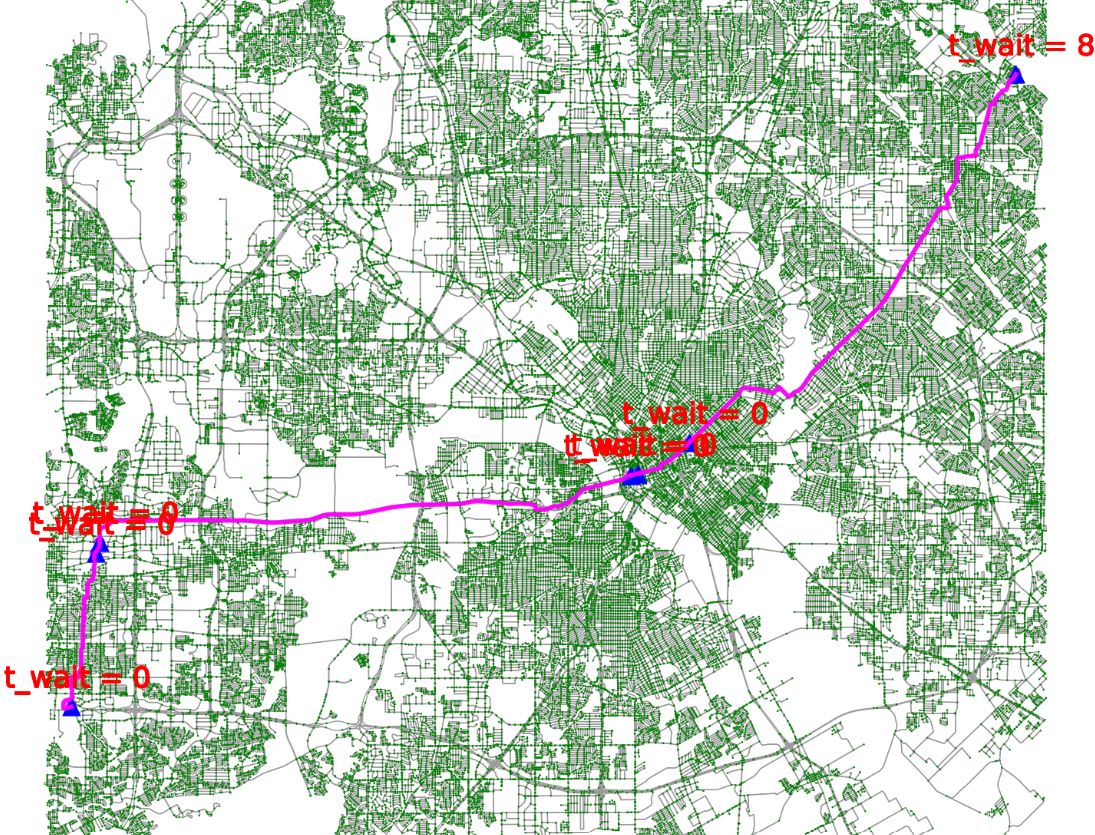}
    \caption{Optimized route planning with minimum waiting and traveling time }
    \label{fig:chargertradeoff_route}
\end{figure}

\section{Conclusion}\label{conc}
In this study, we analyzed the EV charging infrastructure across Texas, with a specific focus on the Dallas-Fort Worth (DFW) area. By integrating effective charging strategies, addressing the sparse distribution of charging stations, and considering the waiting times due to congestion, we have identified a network characterized by significant connectivity and the presence of pivotal hubs. These are essential for network resilience and efficient EV travel within urban areas. Our findings introduce an innovative method for optimizing charger placement that accounts for traffic congestion and the charging efficiency at CC and CV phases. This approach is designed to enhance the practicality of the infrastructure and cater to the specific travel habits of EV users in the DFW area.


By analyzing the current EV charger coverage and suggesting strategic planning, this study contributes to the understanding of infrastructure needs for supporting 
EV travel, emphasizing the importance of charger placement and congestion management in facilitating broader EV adoption.

\section*{Acknowledgments}
This work was partially supported by the Office of Naval Research under ONR award number N00014-21-1-2530 and National Science Foundation under grant 2229417. The United States Government has a royalty-free license throughout the world in all copyrightable material contained herein. Any opinions, findings, and conclusions or recommendations expressed in this material are those of the author(s) and do not necessarily reflect the views of the Office of Naval Research and National Science Foundation.




\bibliography{KPEC.bib}
\vspace{12pt}
\end{document}